\documentclass[conference,twocolumn,10pt]{IEEEtran} 
\usepackage[]{times} 
\usepackage{epsfig} 
\usepackage{subfigure}
\usepackage{amsmath} 
\usepackage{amssymb} 
\usepackage{graphicx} 
\usepackage{subfigure}
\usepackage{multirow} 
\renewcommand{\baselinestretch}{1} 
\newcommand{\bega}{\begin{eqnarray}}
\newcommand{\ega}{\end{eqnarray}}
\newcommand{\bb}{\begin{equation}}
\newcommand{\ee}{\end{equation}}
\newtheorem{defn} {Definition}
\newtheorem{te}{Theorem}
\newtheorem{lema}{Lemma}
\newtheorem{cor}{Corollary}

\begin{document}
\title{On the guaranteed error correction capability of LDPC codes}

\author{\authorblockN{Shashi Kiran Chilappagari, Dung Viet Nguyen, Bane Vasi\'{c}, Michael Marcellin }
\authorblockA{Dept. of Electrical and Computer Eng.\\
University of Arizona\\
Tucson, AZ 85721, USA\\
Email: \{shashic,nguyendv,vasic,marcellin\}@ece.arizona.edu}}

\maketitle
\thispagestyle{empty}
\begin{abstract}
We investigate the relation between the girth and the guaranteed error correction capability of $\gamma$-left regular LDPC codes when decoded using the bit flipping (serial and parallel) algorithms. A lower bound on the number of variable nodes which expand by a factor of at least $3 \gamma/4$ is found based on the Moore bound. An upper bound on the guaranteed correction capability is established by studying the sizes of smallest possible trapping sets. 
\end{abstract}
\section{Introduction}
Iterative algorithms for decoding low-density parity-check (LDPC) codes have been the focus of research over the past decade and most of their properties are well understood \cite{richardsonurbanke,richardsonurbankeshokrollahi}.  These algorithms operate by passing messages along the edges of a graphical representation of the code known as the Tanner graph, and are optimal when the underlying graph is a tree. Message passing decoders perform remarkably well which can be attributed to their ability to correct errors beyond the traditional bounded distance decoding capability. However, in contrast to bounded distance decoders (BDDs), the guaranteed error correction capability of iterative decoders is largely unknown.

The problem of recovering from a fixed number of erasures is solved for iterative decoding on the binary erasure channel (BEC). If the Tanner graph of a code does not contain any stopping sets \cite{di} up to size $t$ (or equivalently the size of minimum stopping set is $t+1$), then the decoder is guaranteed to recover from any $t$ erasures. Orlitsky \textit{et al.} in \cite{orlitsky} studied the relation between stopping sets and girth and derived bounds on the smallest stopping set in any $d$-left regular bipartite graph and girth $g$.

An analogous result is unknown for decoding on other channels such as the binary symmetric channel (BSC) and the additive white Gaussian noise (AWGN) channel. In this paper, we present a step toward such result for hard decision decoding algorithms. Gallager \cite{gallager} proposed two binary message passing algorithms, namely Gallager A and Gallager B, for decoding over the BSC. He showed that for the column-weight $\gamma \geq 3$ and $\rho >\gamma$, there exist $(n,\gamma,\rho)$ \footnote{Precise definitions will be given in Section \ref{section2}} regular low-density parity-check (LDPC) codes for which the bit error probability asymptotically tends to zero  whenever we operate below the threshold. The minimum distance was shown to increase linearly with the code length, but correction of a linear fraction of errors was not shown. Zyablov and Pinsker \cite{zyablov} analyzed LDPC codes under a simpler decoding algorithm known as the bit flipping algorithm and showed that almost all the codes in the regular ensemble with $\gamma \geq 5$ can correct a constant fraction of worst case errors. Sipser and Spielman in \cite{spielman} used expander graph arguments to analyze two bit flipping algorithms, serial and parallel. Specifically, they showed that these algorithms can correct a fraction of errors if the underlying Tanner graph is a good expander. Burshtein and Miller in \cite{burshtein} applied expander based arguments to show that message passing algorithms can also correct a fixed fraction of worst case errors when the degree of each variable node is more than five. Feldman \textit{et al.} \cite{feldman} showed that the linear programming decoder \cite{feldman2} is also capable of correcting a fraction of errors. Recently, Burshtein in \cite{burshteinisitpaper} showed that regular codes with variable nodes of degree four  are capable of correcting a linear number of errors under bit flipping algorithm. He also showed tremendous improvement in the fraction of correctable errors when the variable node degree  is at least five.

It is well known that a random graph is a good expander with high probability \cite{spielman}. However, the fraction of nodes having the required expansion is very small and hence the code length to guarantee correction of a fixed number of errors must be large. Moreover, determining the expansion of a given graph is known to be NP hard \cite{alon}, and spectral gap methods cannot guarantee an expansion factor of more than 1/2 \cite{spielman}. On the other hand, code parameters such as column weight and girth can be easily determined or are assumed to be known for the code under consideration. The approach in this paper is to determine the size of variable node sets in a left regular LDPC code which are guaranteed to have the expansion required by bit flipping algorithms based on Moore bound \cite[p.180]{biggs}. The consequence of our results is that the error correction capability grows exponentially in girth. However, we note that since the girth grows logarithmically in the code length, this result does not show that the bit flipping algorithms can correct a linear fraction of errors (the proof and discussion are beyond the scope of this paper).    

To find an upper bound on the number of correctable errors, we study the size of sets of variable nodes which lead to decoding failures. A decoding failure is said to have occurred if the output of the decoder is not equal to the transmitted codeword \cite{rich}. The conditions that lead to decoding failures are well understood for a variety of decoding algorithms such as maximum likelihood decoding, bounded distance decoding and iterative decoding on the binary erasure channel. However, for iterative decoding on the BSC and AWGN channel, the understanding is far from complete. Two approaches have been taken in this direction, namely trapping sets \cite{rich} and pseudo-codewords \cite{koetter}. We adopt the trapping set approach in this paper to characterize decoding failures. Richardson introduced the notion of trapping sets in \cite{rich} to estimate the error floor on the AWGN channel. In \cite{chilappagarione}, trapping sets were used to estimate the frame error rate of column-weigh-three LDPC codes.  In this paper, we define trapping sets with the help of fixed points for the bit flipping algorithms (both serial and parallel). We then find bounds on the size of trapping sets based on extremal graphs known as cage graphs \cite{cage}, thereby finding an upper bound on the guaranteed error correction capability. 

The rest of the paper is organized as follows. In Section \ref{section2}, we provide a brief introduction to LDPC codes, decoding algorithms and trapping sets \cite{rich}. In Section \ref{section3}, we prove our main theorem relating the column weight and girth to the number of variable nodes which expand by a factor of at least $3 \gamma/4$. We derive bounds on the size of trapping sets in Section \ref{section4}, and conclude with a few remarks in Section \ref{section5}. 
\section{Preliminaries}\label{section2}
In this section, we first establish the notation and then proceed to give a brief introduction to LDPC codes and bit flipping algorithms. We then give the relation between the error correction capability of the code and the expansion of the underlying Tanner graph. We finally describe trapping sets for the bit flipping algorithms. 
\subsection{Graph Theory Notation}
We adopt the standard notation in graph theory (see \cite{bollobas} for example).
$G=(U,E)$ denotes a graph with set of nodes $U$ and set of edges $E$. When there is no ambiguity, we simply denote the graph by $G$. An edge $e$ is an unordered pair $(u_1,u_2)$ of nodes and is said to be incident on $u_1$ and $u_2$. Two nodes $u_1$ and $u_2$ are said to be adjacent (neighbors) if there is an edge $e=(u_1,u_2)$ incident on them. The order of the graph is $|U|$ and the size of the graph is $|E|$. The degree of $u$, $d(u)$, is the number of its neighbors. A node with degree one is called a leaf or a pendant node. A graph is $d$-regular if all the nodes have degree $d$. The average degree $\overline{d}$ of a graph is defined as $\overline{d}=2|E|/|U|$. The girth $g(G)$ of a graph $G$, is the length of smallest cycle in $G$. $H=(V \cup C,E')$ denotes a bipartite graph with two sets of nodes; variable (left) nodes $V$ and check (right) nodes $C$ and edge set $E'$. Nodes in $V$ have neighbors only in $C$ and vice versa. A bipartite graph is said to be $\gamma$-left regular if all variable nodes have degree $\gamma$, $\rho$-right regular if all check nodes have degree $\rho$ and $(\gamma,\rho)$ regular if all variable nodes have degree $\gamma$ and all check nodes have degree $\rho$. The girth of a bipartite graph is even. 
\subsection{LDPC Codes and Decoding Algorithms}
LDPC codes \cite{gallager} are a class of linear block codes which can be defined by sparse bipartite graphs \cite{shokrollahi}. Let $G$ be a bipartite graph with two sets of nodes: $n$ variable nodes and $m$ check nodes. This graph defines a linear block code $\mathcal{C}$ of length $n$ and dimension at least $n-m$ in the following way: The $n$ variable nodes are associated to the $n$ coordinates of codewords. A vector $\mathbf{v}=(v_1,v_2,\ldots,v_n)$ is a codeword if and only if for each check node, the modulo two sum  of its neighbors is zero.  Such a graphical representation of an LDPC code is called the Tanner graph \cite{tanner} of the code. The adjacency matrix of $G$ gives a parity check matrix of $\cal{C}$. An $(n,\gamma,\rho)$ regular LDPC code has a Tanner graph with $n$ variable nodes each of degree  $\gamma$ (column weight) and $n\gamma/ \rho$ check nodes each of degree  $\rho$ (row weight). This code has length $n$ and rate $r \geq 1-\gamma/\rho$ \cite{shokrollahi}.

We now describe the parallel bit flipping algorithm \cite{zyablov,spielman} to decode LDPC codes. As noted earlier, each check node imposes a constraint on the neighboring variable nodes. A constraint (check node) is said to be satisfied by a setting of variable nodes if the sum of the variable nodes in the constraint is even; otherwise the constraint is unsatisfied.

{\bfseries Parallel Bit Flipping Algorithm}
\begin{itemize}
\item In parallel, flip each variable that is in more unsatisfied than satisfied constraints.
\item Repeat until no such variable remains. 
\end{itemize}
A serial version of the algorithm is also defined in \cite{spielman} and all the results in this paper hold for the serial bit flipping algorithm also. The bit flipping algorithms are iterative in nature but do not belong to the class of message passing algorithms.
\subsection{Expansion and Error Correction Capability}
Sipser and Spielman \cite{spielman} analyzed the performance of the  bit flipping algorithms using the expansion properties of the underlying Tanner graph of the code. We summarize the results from \cite{spielman} below for the sake of completeness. We start with the following definitions from \cite{spielman}.

\begin{defn}
Let $G=(U,E)$ with $|U|=n_1$. We say that \textit{every set of at most $m_1$ nodes expands by a factor of $\delta$} if, for all sets $S \subset U$
\[
|S|\leq m_1 \Rightarrow |\{y: \exists x \in S \mbox{~such that~} (x,y) \in E \}| > \delta |S|.
\]
\end{defn}
We consider bipartite graphs and expansion of variable nodes only. 
\begin{defn}
A graph is a $(\gamma,\rho,\alpha,\delta)$ expander if it is a $(\gamma,\rho)$ regular bipartite graph in which every subset of at most $\alpha$ fraction of the variable nodes expands by a factor of at least $\delta$.
\end{defn}
The following theorem from \cite{spielman} relates the expansion and error correction capability of an $(n,\gamma,\rho)$ LDPC code with Tanner graph $G$ when decoded using the parallel bit flipping decoding algorithm.
\begin{te}\cite[Theorem 11]{spielman}
Let $G$ be a $(\gamma, \rho, \alpha, (3/4 +\epsilon)\gamma)$ expander over $n$ variable nodes, for any $\epsilon > 0$. Then, the simple parallel decoding algorithm will correct any $\alpha_0 < \alpha(1 + 4\epsilon)/2$ fraction of errors after $\log_{1-4\epsilon}(\alpha_0 n)$ decoding rounds. 
\end{te}
\textit{Notes:}
\begin{enumerate}
\item The serial bit flipping algorithm can also correct $\alpha_0 < \alpha/2$ fraction of errors if $G$ is a $(\gamma, \rho, \alpha, (3/4)\gamma)$ expander.
\item The results hold for any left regular code as we need expansion of variable nodes only.
\end{enumerate}
From the above discussion, we observe that finding the number of variable nodes which are guaranteed to expand by a factor of at least $3 \gamma/4$, gives a lower bound on the guaranteed error correction capability of LDPC codes.
\subsection{Decoding Failures and Trapping Sets}
We now define decoding failures of the bit flipping algorithms and characterize these failures using trapping sets. 

Consider an LDPC code of length $n$ and let $\mathbf{x}$ be the binary vector which is the input to the hard decision decoder. Let $\mathcal{S}(\mathbf{x})$ be the support of $\mathbf{x}$. The support of $\mathbf{x}$ is defined as the set of all positions $i$ where $\mathbf{x}(i) \neq 0$. The set of variable nodes (bits) which differ from their original value are referred to as corrupt variables.
\begin{defn}
$\mathbf{x}$ is a fixed point of the bit flipping algorithm if the set of corrupt variables remains unchanged after one round of decoding.
\end{defn}

\begin{defn}\cite{chilappagarione}
Let $\mathbf{x}$ be a fixed point. Then $\mathcal{S}(\mathbf{x})$ is known as a trapping set. An $(a,b)$ trapping set $\cal{T}$ is a set of $a$ variable nodes whose induced subgraph has $b$ odd degree checks. 
\end{defn}
The size of $\mathcal{T}$ denoted by $|\mathcal{T}|$ is the number of variable nodes in $\mathcal{T}$. From the definitions, it is clear that if the input to the decoder is a fixed point then the output is also the same fixed point. Or in other words, if the initial errors are in variable nodes corresponding to a trapping set, then the decoder will not converge to the original codeword. Hence, bounding the size of trapping sets gives an upper bound on the guaranteed error correction capability. Apart from fixed point decoding failures, there exist other types of failures, the discussion of which is beyond the scope of this paper.
\section{Expansion, Column Weight and Girth}\label{section3}
In this section, we prove our main theorem which relates the column weight and girth of a code to its error correction capability. We show that the size of variable node sets which have the required expansion is related to the well known Moore bound \cite[p.180]{biggs}. We start with a few definitions required to establish the main theorem.
\subsection{Definitions}
\begin{defn}The \textit{reduced graph} $H_r=(V \cup C_r, E'_r)$ of $H=(V \cup C,E')$ is a graph with vertex set $V \cup C_r$ and edge set $E'_r$ given by 
\begin{eqnarray}
C_r &=& C \setminus C_p, ~C_p =\{c \in C : \mbox{c is a pendant node}\} \nonumber \\
E'_r&=& E' \setminus E'_p,~ E'_p = \{(v_i,c_j) \in E : c_j \in C_p\} \nonumber 
\end{eqnarray}
\end{defn}
\begin{defn} Let $H=(V \cup C, E')$ be such that $\forall v \in V, d(v) \leq \gamma$. The \textit{$\gamma$ augmented graph} $H_{\gamma}=(V \cup C_{\gamma}, E'_{\gamma})$ is a graph with vertex set $V \cup C_{\gamma}$ and edge set $E'_{\gamma}$ given by
\begin{eqnarray}
C_{\gamma} &=& C \cup C_a, \mbox{~where~} C_a = \bigcup_{i=1}^{|V|}C_a^i \mbox{~and~} \nonumber \\
C_a^i &=& \{c_1^i,\ldots,c_{\gamma-d(v_i)}^i\}. \nonumber \\
E'_{\gamma}&=& E' \cup E'_{a}, \mbox{~where~} E'_a = \bigcup_{i=1}^{|V|} E_{a}^{'i} \mbox{~and} \nonumber \\
E_a^{'i} &=& \{(v_i,c_j)\in V \times C_{a}: c_j \in C_a^i\}. \nonumber
\end{eqnarray}
\end{defn}
  
\begin{defn}\cite[Definition 4]{spielman} The \textit{edge-vertex incidence graph} $G_{ev}=(U \cup E, E_{ev})$ of $G=(U,E)$ is the bipartite graph with vertex set $U \cup E$  and edge set
\[
E_{ev}=\{(e,u) \in E \times U : \mbox{$u$ is an endpoint of e}\}.
\]\end{defn}
\textit{Notes:}  
\begin{enumerate}
\item The edge-vertex incidence graph is right regular with degree two.
\item $|E_{ev}|=2|E|$.
\item $g(G_{ev})=2g(G)$.
\end{enumerate}

\begin{defn}An \textit{inverse edge-vertex incidence graph} $H_{iev}=(V, E'_{iev})$ of $H=(V \cup C, E')$ is a graph with vertex set $V$ and edge set $E'_{iev}$ which is obtained as follows. For $c \in C_r$, let $N(c)$ denote the set of neighbors of $c$. Label one node $v_i \in N(c)$ as a root node. Then 
\begin{eqnarray}
E'_{iev}&=&\{(v_i,v_j) \in V \times V: v_i \in N(c), v_j \in N(c),  \nonumber \\
&&i \neq j,\mbox { $v_i$ is a root node, for some $c \in C_r$} \}. \nonumber
\end{eqnarray}
\end{defn}
\textit{Notes:}  
\begin{enumerate}
\item Given a graph, the inverse edge-vertex incidence graph is not unique. 
\item $g(H_{iev}) \geq g(H)/2$, $|E'_{iev}| = |E'_r| - |C_r|$ and $|C_r| \leq |E'_r|/2$
\item $|E'_{iev}| \geq |E'_r|/2$ with equality only if all checks in $C_r$ have degree two.
\item The term inverse edge-vertex incidence is used for the following reason. Suppose all checks in $H$ have degree two. Then the edge-vertex incidence graph of $H_{iev}$ is $H$.
\end{enumerate}

The \textit{Moore bound} \cite[p.180]{biggs} denoted by $n_0(d,g)$ is a lower bound on the least number of vertices in a $d$-regular graph with girth $g$. It is given by 
\begin{eqnarray}
n_0(d,g)=n_0(d,2r+1) &=& 1 + d \sum_{i=0}^{r-1} (d-1)^i, g~\mbox{odd} \nonumber\\
n_0(d,g)=n_0(d,2r)&=& 2 \sum_{i=0}^{r-1}(d-1)^i \nonumber, g~\mbox{even}
\end{eqnarray}

In \cite{mooreirreg}, it was shown that a similar bound holds for irregular graphs. 
\begin{te}\cite{mooreirreg}
The number of nodes $n(\overline{d},g)$ in a graph of girth $g$ and average degree at least $\overline{d} \geq 2$ satisfies:
\[
n(\overline{d},g) \geq n_0(\overline{d},g)
\]
\end{te}
Note that $\overline{d}$ need not be an integer in the above theorem.

\subsection{The Main Theorem}
We now state and prove the main theorem.
\begin{te}Let $\mathcal{C}$ be an LDPC code with $\gamma$-left regular Tanner graph $G$. Let $g(G)=2g'$. Then for all $k < n_0(\gamma/2,g')$, any set of $k$ variable nodes expands by a factor of at least $3 \gamma/4$.\end{te}
\begin{proof}
Let $G^{k}=(V^k \cup C^k, E^k )$ denote the subgraph induced by a set of $k$ variable nodes $V^{k}$. Since $G$ is $\gamma$-left regular, $|E^{k}|=\gamma k$. Let $G^{k}_r=(V^{k} \cup C^{k}_r ,E^{k}_r)$ be the reduced graph. We have
\begin{eqnarray}
|C^{k}| &=& |C^{k}_r| + |C^{k}_p| \nonumber \\
|E^k| &=& |E^k_p| + |E^k_r| \nonumber \\
|E^k_p| &=& |C^{k}_p| \nonumber \\ 
|C^{k}_p| &=& \gamma k - |E^{k}_r| \nonumber 
\end{eqnarray}
We need to prove that $|C^k| > 3\gamma k/4$. 

Let $f(k,g')$ denote the maximum number of edges in an arbitrary graph of order $k$ and girth $g'$. By Theorem 2, for all $k < n_0 (\gamma/2,g')$, the average degree of a graph with $k$ nodes and girth $g'$ is less than $\gamma/2$. Hence, $f(k,g') < \gamma k/4$.  We now have the following lemma.

\begin{lema}
The number of edges in  $G^{k}_r$ cannot exceed $2f(k,g')$ i.e.,
\[
|E^{k}_r| \leq  2 f(k,g').
\]
\end{lema}
\begin{proof}
The proof is by contradiction. Assume that $|E^{k}_r| > 2f(k,g')$. Consider $G^{k}_{iev}=(V^{k}, E^{k}_{iev})$, an inverse edge vertex incidence graph of $G^{k}$. We have
\[
|E^{k}_{iev}| > f(k,g'). 
\]
This is a contradiction as $G^{k}_{eiv}$ is a graph of order $k$ and girth at least $g'$.
\end{proof}
We now find a lower bound on $|C^k|$  in terms of $f(k,g')$. We have the following lemma.
\begin{lema}
$|C^{k}| \geq \gamma k - f(k,g')$. 
\end{lema}
\begin{proof}
Let $|E^{k}_{r}| = 2f(k,g') - j$ for some integer $j \geq 0$. Then $|E^{k}_{p}| = \gamma k - 2f(k,g') + j$. We claim that  $|C^{k}_{r}| \geq f(k,g') + j$. To see this, we note that 
\begin{eqnarray}
|E^{k}_{iev}| &=& |E^{k}_{r}| - |C^{k}_{r}|, \mbox{~or} \nonumber \\
|C^{k}_{r}| &=& |E^{k}_{r}| - |E^{k}_{iev}| \nonumber 
\end{eqnarray}
But
\begin{eqnarray}
|E^{k}_{iev}| &\leq& f(k,g') \nonumber \\
\Rightarrow |C^{k}_{r}| &\geq& 2f(k,g') - j - f(k,g') \nonumber \\
\Rightarrow |C^{k}_{r}| &\geq& f(k,g') - j \nonumber
\end{eqnarray}
Hence we have,
\begin{eqnarray}
|C^{k}| &=&  |C^{k}_{r}| + |C^{k}_{p}| \nonumber \\
\Rightarrow |C^{k}| &\geq& f(k,g') - j + \gamma k - 2f(k,g') + j \nonumber \\
\Rightarrow |C^{k}| &\geq& \gamma k - f(k,g'). \nonumber
\end{eqnarray}
\end{proof}
The theorem now follows as
\[
f(k,g') < \gamma k/4 
\]
and therefore
\[
|C^{k}| > 3\gamma k/4
\]
\end{proof}
\begin{cor}
Let $\mathcal{C}$ be an LDPC code with column-weight $\gamma$ and girth $2g'$. Then the bit flipping algorithm can correct any error pattern of weight less than $n_0(\gamma/2,g')/2$.
\end{cor}
\section{Cage Graphs and Trapping Sets}\label{section4}
In this section, we first give necessary and sufficient conditions for a given set of variables to be a trapping set. We then proceed to define a class of interesting graphs known as cage graphs and establish a relation between cage graphs\cite{cage} and trapping sets. We then give an upper bound on the error correction capability based on the sizes of cage graphs. The proofs in this section are along the same lines as in Section \ref{section3}. Hence, due to space considerations, we only give a sketch of the proofs.
\begin{te}\label{thm1}
Let $\mathcal{C}$ be an LDPC code with $\gamma$-left regular Tanner graph $G$. Let $\cal{T}$ be a set consisting of $V$ variable nodes with induced subgraph $\cal{I}$. Let the checks in $\cal{I}$ be partitioned into two disjoint subsets; $\cal{O}$ consisting of checks with odd degree and $\cal{E}$ consisting of checks with even degree. Then $\cal{T}$ is a trapping set for bit flipping algorithm iff : (a) Every variable node in $\cal{I}$ has at least $\left\lceil \gamma/2 \right\rceil$  neighbors in $\cal{E}$, and (b) No $\left\lfloor \gamma/2 \right\rfloor + 1$ checks of $\cal{O}$ share a  neighbor outside $\cal{I}$.
\end{te}
\begin{proof}
We first show that the conditions stated are sufficient. Let $\mathbf{x_{\mathcal{T}}}$ be the input to the bit flipping algorithm, with support $\mathcal{T}$. The only unsatisfied constraints are in $\mathcal{O}$. By the conditions of the theorem, we observe that no variable node is involved in more unsatisfied constraints than satisfied constraints. Hence, no variable node is flipped and by definition  $\mathbf{x_{\mathcal{T}}}$ is a fixed point implying that $\mathcal{T}$ is a trapping set.

To see that the conditions are necessary, observe that for $\mathbf{x}_{\mathcal{T}}$ to be a trapping set, no variable node should be involved in more unsatisfied constraints than satisfied constraints. 
\end{proof}

\textit{Remark:} Theorem \ref{thm1} is a consequence of Fact 3 from \cite{rich}.

To determine whether a given set of variables is a trapping set, it is necessary to not only know the induced subgraph but also the neighbors of the odd degree checks. However, in order to establish general bounds on the sizes of trapping sets given only the column weight and the girth, we consider only condition (a) of Theorem \ref{thm1} which is a necessary condition. A set of variable nodes satisfying condition (a) is known as a \textit{potential trapping set}. A trapping set is a potential trapping set that satisfies condition (b). Hence, finding bounds on the size of potential trapping sets gives bounds on the size of trapping sets. 
\begin{defn}\cite{cage}
A $(d,g)$-\textit{cage graph}, $G(d,g)$, is a $d$-regular graph with girth $g$ having the minimum possible number of nodes.
\end{defn}
A lower bound, $n_l(d,g)$, on the number of nodes $n_c(d,g)$ in a $(d,g)$-cage graph is given by the Moore bound. An upper bound $n_u(d,g)$ on $n_c(d,g)$ (see \cite{cage} and references therein) is given by
\begin{eqnarray}
n_u(3,g)&=& \left\{\begin{array}{cl}\frac{4}{3} + \frac{29}{12}~2^{g-2} & \mbox{for g odd} \\
																	\frac{2}{3} + \frac{29}{12}~2^{g-2} & \mbox{for g even} \end{array} \right. \nonumber \\
n_u(d,g)&=& \left\{\begin{array}{cl} 2(d-1)^{g-2} & \mbox{for g odd} \\
																	 4(d-1)^{g-3}& \mbox{for g even} \end{array} \right. \nonumber
\end{eqnarray}
\begin{te}
Let $\mathcal{C}$ be an LDPC code with $\gamma$-left regular Tanner graph $G$ and girth $2g'$. Let $\mathcal{T}(\gamma,2g')$ denote the the smallest possible trapping set of $\mathcal{C}$ for the bit flipping algorithm. Then,
\[
|\mathcal{T}(\gamma,2g')| = n_c(\left\lceil \gamma/2 \right\rceil,g')
\]
\end{te}
\begin{proof}
We first find a lower bound on $|\mathcal{T}(\gamma,2g')|$ and then exhibit a potential trapping set of size $n_c(\left\lceil \gamma/2 \right\rceil,g')$. We begin with the following lemma.
\begin{lema}
$|\mathcal{T}(\gamma,2g')| \geq n_c(\left\lceil \gamma/2 \right\rceil,g')$.
\end{lema}
\begin{proof}
Let $\mathcal{T}_1$ be a trapping set with $|\mathcal{T}_1|< n_c(\left\lceil \gamma/2 \right\rceil,g')$ and let $G_1$ denote the induced subgraph of $\mathcal{T}_1$. We can construct a $(\left\lceil \gamma/2 \right\rceil,g'')$- cage graph $(g'' \geq g)$ with $|\mathcal{T}_1|< n_c(\left\lceil \gamma/2 \right\rceil,g')$ nodes by removing edges (if necessary) from the inverse edge-vertex of $G_1$ which is a contradiction.
\end{proof}
We now exhibit a potential trapping set of size $n_c(\left\lceil \gamma/2 \right\rceil,g')$. Let $G_{ev}(\left\lceil \gamma/2 \right\rceil,g')$ be the edge-vertex incidence graph of a $G(\left\lceil \gamma/2 \right\rceil,g')$. Note that $G_{ev}(\left\lceil \gamma/2 \right\rceil,g')$ is a left regular bipartite graph with $n_c(\left\lceil \gamma/2 \right\rceil,g')$ variable nodes of degree $\left\lceil \gamma/2 \right\rceil$ and all checks have degree two. Now consider $G_{ev,\gamma}(\left\lceil \gamma/2 \right\rceil,g')$, the $\gamma$ augmented graph of $G_{ev}(\left\lceil \gamma/2 \right\rceil,g')$. It can be seen that $G_{ev,\gamma}(\left\lceil \gamma/2 \right\rceil,g')$ is a potential trapping set.
\end{proof}
\begin{cor}
Let $\mathcal{C}$ be an LDPC code with column-weight $\gamma$ and girth $2g'$. Then the bit flipping algorithm cannot be guaranteed to correct all error patterns of weight more than or equal to $n_c(\left\lceil \gamma/2 \right\rceil,g')$.
\end{cor}
\section{Discussion}\label{section5}
We derived lower bounds and upper bounds on the guaranteed error correction capability of left regular LDPC codes. The lower bounds we derived in this paper are weak. However, extremal graphs avoiding three, four and five cycles have been studied in great detail (see \cite{extremalone,extremaltwo}) and these results can be used to derive tighter bounds when the girth is eight, ten or twelve. Also, since an expansion factor of $3 \gamma/4$ is not necessary (see \cite[Theorem 24]{spielman}), it is possible that tighter lower bounds can be derived for some cases. The results can be extended to Gallager A and Gallager B algorithms as well. It should be noted that the necessary and sufficient conditions for a set to be trapping set for Gallager A/B algorithms are similar (depending on the message passing rules) to those in Theorem \ref{thm1}. Our approach can be used to derive bounds on the guaranteed erasure recovery capability for iterative decoding on the BEC by finding number of variable nodes which expand by a factor of $\gamma/2$. In \cite{orlitsky}, the bounds on the guaranteed erasure recovery capability were derived based on the size of smallest stopping set. Both approaches give the same bound, which also coincide with the bounds given by Tanner in \cite{tanner} for the minimum distance. 
\section*{Acknowledgment}
This work is funded by NSF under Grant CCF-0634969, ITR-0325979, ECCS-0725405 and INSIC-EHDR program.


\begin{thebibliography}{10}
\providecommand{\url}[1]{#1}
\csname url@rmstyle\endcsname
\providecommand{\newblock}{\relax}
\providecommand{\bibinfo}[2]{#2}
\providecommand\BIBentrySTDinterwordspacing{\spaceskip=0pt\relax}
\providecommand\BIBentryALTinterwordstretchfactor{4}
\providecommand\BIBentryALTinterwordspacing{\spaceskip=\fontdimen2\font plus
\BIBentryALTinterwordstretchfactor\fontdimen3\font minus
  \fontdimen4\font\relax}
\providecommand\BIBforeignlanguage[2]{{%
\expandafter\ifx\csname l@#1\endcsname\relax
\typeout{** WARNING: IEEEtran.bst: No hyphenation pattern has been}%
\typeout{** loaded for the language `#1'. Using the pattern for}%
\typeout{** the default language instead.}%
\else
\language=\csname l@#1\endcsname
\fi
#2}}

\bibitem{richardsonurbanke}
T.~J. Richardson and R.~Urbanke, ``The capacity of low-density parity-check
  codes under message-passing decoding,'' \emph{IEEE Trans. Inform. Theory},
  vol.~47, no.~2, pp. 599--618, Feb. 2001.

\bibitem{richardsonurbankeshokrollahi}
T.~J. Richardson, M.~Shokrollahi, and R.~Urbanke, ``Design of
  capacity-approaching irregular low-density parity-check codes,'' \emph{IEEE
  Trans. Inform. Theory}, vol.~47, no.~2, pp. 638--656, Feb. 2001.

\bibitem{di}
C.~Di, D.~Proietti, T.~Richardson, E.~Telatar, and R.~Urbanke, ``Finite length
  analysis of low-density parity-check codes,'' \emph{IEEE Trans. Inform.
  Theory}, vol.~48, pp. 1570--1579, June 2002.

\bibitem{orlitsky}
A.~Orlitsky, R.~Urbanke, K.~Viswanathan, and J.~Zhang, ``Stopping sets and the
  girth of tanner graphs,'' in \emph{Proc. of IEEE International Symposium on
  Information Theory}, 2002, p.~2.

\bibitem{gallager}
R.~G. Gallager, \emph{Low Density Parity Check Codes}.\hskip 1em plus 0.5em
  minus 0.4em\relax Cambridge, MA: M.I.T. Press, 1963.

\bibitem{zyablov}
V.~V. Zyablov and M.~S. Pinsker, ``Estimation of the error-correction
  complexity for {G}allager low-density codes,'' \emph{Problems of Information
  Transmission}, vol.~11, no.~1, pp. 18--28, 1976.

\bibitem{spielman}
M.~Sipser and D.~Spielman, ``Expander codes,'' \emph{IEEE Trans. Inform.
  Theory}, vol.~42, no.~6, pp. 1710--1722, Nov. 1996.

\bibitem{burshtein}
D.~Burshtein and G.~Miller, ``Expander graph arguments for message-passing
  algorithms,'' \emph{IEEE Trans. Inform. Theory}, vol.~47, no.~2, pp.
  782--790, Feb. 2001.

\bibitem{feldman}
J.~Feldman, T.~Malkin, R.~A. Servedio, C.~Stein, and M.~J. Wainwright, ``L{P}
  decoding corrects a constant fraction of errors,'' \emph{IEEE Trans. Inform.
  Theory}, vol.~53, no.~1, pp. 82--89, Jan. 2007.

\bibitem{feldman2}
J.~Feldman, M.~J. Wainwright, and D.~R. Karger, ``Using linear programming to
  decode binary linear codes,'' \emph{IEEE Trans. Inform. Theory}, vol.~51,
  no.~3, pp. 954--972, March 2005.

\bibitem{burshteinisitpaper}
D.~Burshtein, ``On the error correction of regular {LDPC} codes using the
  flipping algorithm,'' in \emph{Proc. of IEEE International Symposium on
  Information Theory}, June 2007, pp. 226--230.

\bibitem{alon}
N.~Alon, ``Spectral techniques in graph algorithms,'' in \emph{LATIN '98:
  Proceedings of the Third Latin American Symposium on Theoretical
  Informatics}.\hskip 1em plus 0.5em minus 0.4em\relax London, UK:
  Springer-Verlag, 1998, pp. 206--215.

\bibitem{biggs}
N.~Biggs, \emph{Algebraic graph theory}.\hskip 1em plus 0.5em minus 0.4em\relax
  Cambridge: Cambridge University Press, 1993.

\bibitem{rich}
T.~J. Richardson, ``Error floors of {LDPC} codes,'' in \emph{Proc. of 41st
  Annual Allerton Conf. on Communications, Control and Computing}, 2003, pp.
  1426--1435.

\bibitem{koetter}
\BIBentryALTinterwordspacing
P.~O. Vontobel and R.~Koetter, ``Graph-cover decoding and finite-length
  analysis of message-passing iterative decoding of {LDPC} codes,'' May 2007,
  accepted for IEEE Transactions on Information Theory. [Online]. Available:
  \url{http://www.citebase.org/abstract?id=oai:arXiv.org:cs/0512078}
\BIBentrySTDinterwordspacing

\bibitem{chilappagarione}
S.~K. Chilappagari, S.~Sankaranarayanan, and B.~Vasic, ``Error floors of {LDPC}
  codes on the binary symmetric channel,'' in \emph{Proc. of IEEE International
  Conference on Communications}, vol.~3, June 11-15 2006, pp. 1089--1094.

\bibitem{cage}
\BIBentryALTinterwordspacing
E.~W. Weisstein, ``Cage graph.'' [Online]. Available:
  \url{http://mathworld.wolfram.com/CageGraph.html}
\BIBentrySTDinterwordspacing

\bibitem{bollobas}
B.~Bollobas, \emph{Extremal graph theory}.\hskip 1em plus 0.5em minus
  0.4em\relax London: Academic Press Inc., 1978.

\bibitem{shokrollahi}
A.~Shokrollahi, ``An introduction to low-density parity-check codes,'' in
  \emph{Theoretical aspects of computer science: advanced lectures}.\hskip 1em
  plus 0.5em minus 0.4em\relax New York, NY, USA: Springer-Verlag New York,
  Inc., 2002, pp. 175--197.

\bibitem{tanner}
R.~M. Tanner, ``A recursive approach to low complexity codes,'' \emph{IEEE
  Trans. Inform. Theory}, vol.~27, no.~5, pp. 533--547, Sept. 1981.

\bibitem{mooreirreg}
N.~Alon, S.~Hoory, and M.~Linial, ``The moore bound for irregular graphs,''
  \emph{Graphs and Combinatorics}, vol.~18, no.~1, pp. 53--57, 2002.

\bibitem{extremalone}
D.~K. Garnick, Y.~H.~H. Kwong, and F.~Lazebnik, ``Extremal graphs without
  three-cycles or four-cycles,'' \emph{J. Graph Theory}, vol.~17, no.~5, pp.
  633--645, 1993.

\bibitem{extremaltwo}
Y.~Yuansheng, L.~Xiaohui, D.~Guocheng, and Z.~Yongxiang, ``Extremal graphs
  without three-cycles, four-cycles or five-cycles,'' \emph{Utilitas
  Mathematica}, vol.~66, pp. 249--266, 2004.

\end{thebibliography}
\end{document}